\documentclass[conference]{IEEEtran}
\IEEEoverridecommandlockouts
\usepackage{cite}
\usepackage{amsmath,amssymb,amsfonts}
\usepackage{algorithmic}
\usepackage{graphicx}
\usepackage{textcomp}
\usepackage{xcolor}
\usepackage{booktabs}
\usepackage{multirow}
\usepackage{makecell}

\def\BibTeX{{\rm B\kern-.05em{\sc i\kern-.025em b}\kern-.08em
    T\kern-.1667em\lower.7ex\hbox{E}\kern-.125emX}}
\begin{document}

\title{DRGM: Data Requirement Goal Modeling for ML-Based Systems}

\author{
Asma Yamani\textsuperscript{1},
Nadeen AlAmoudi\textsuperscript{1},
Salma Albelali\textsuperscript{1},
Malak Baslyman\textsuperscript{1,2},
Jameleddine Hassine\textsuperscript{1,3} \\
\textsuperscript{1}Information and Computer Science Department, KFUPM, Dhahran, Saudi Arabia \\
\textsuperscript{2}IRC for Finance and Digital Economy, KFUPM, Dhahran, Saudi Arabia \\
\textsuperscript{3}IRC for Intelligent Secure Systems, KFUPM, Dhahran, Saudi Arabia \\
\{g201906630, g201906430, g201907430, malak.baslyman, jhassine\}@kfupm.edu.sa
}


\maketitle

\begin{abstract}
Given the critical role of data in Machine Learning (ML)-based system development, it has become increasingly important to assess the quality of data attributes and ensure that the data meets specific requirements before its utilization. This work proposes an approach to guide non-experts in identifying data requirements for ML systems using goal modeling. In this approach, we first develop the Data Requirement Goal Model (DRGM) by surveying the scientific literature to identify and categorize the issues and challenges faced by data scientists and requirement engineers working on ML-related projects. An initial DRGM was built to accommodate common tasks that would generalize across projects. Then, a customization mechanism is built to help adjust the tasks, KPIs, and goals' importance of different elements within the DRGM. The generated model can aid its users in evaluating different datasets using Goal-oriented Requirement Language (GRL) evaluation strategies. We then validate the approach through two illustrative examples based on real-world projects. The results demonstrate that the data requirements identified by the proposed approach align with the requirements of real-world projects, showing the practicality and effectiveness of the proposed framework. For future work, we recommend further evaluation of the proposed approach across more ML problem types and contexts, as well as implementing tool support for generating the DRGM via a chatbot interface.

\end{abstract}

\begin{IEEEkeywords}
Goal model, data requirements, data quality
\end{IEEEkeywords}
\section{Introduction}
\label{section1}
Machine Learning (ML) is being integrated into various real-life applications, including medical diagnosis, stock market trading, and image recognition applications necessitating an investigation into how ML is addressed during the various stages of software development~\cite{r1,r2,r3,r4,r7,r6,r5}. In contrast to conventional software systems, the behavior of ML-based systems is primarily driven by data and the model constructed from it rather than by predefined functionalities and logic designed by engineers~\cite{r1,r4,r6,r2}. Consequently, in the realm of ML, data requirements and properties exert a profound influence on model performance, interpretability, and fairness. Data requirements should play a pivotal role in guiding feature selection, preprocessing, and ensuring that the data is effective and safe for learning and prediction.\par Understanding data properties, such as distribution and correlations, is essential for model interpretability and significantly impacts model evaluation and validation. Addressing bias and fairness in ML models depends on a comprehensive understanding of data properties to effectively mitigate biases. One example that highlights the importance of data properties to building an effective ML system, is a systemic literature review investigating the clinical viability of machine learning models developed to detect and diagnose COVID-19 from chest x-rays in studies published in 2020. This review found that none of the included models demonstrated potential clinical utility~\cite{Roberts2021}. This was due to methodological flaws or underlying biases, related to the datasets issues, including the use of public datasets where the integrity of the data is questioned, challenges with training data size, balancedness, and the presence of the control group. Another example is the case of the Amazon INC recruiting ML system. Although the gender of the applicant was protected, the models were able to recognize patterns in women's writing style, which led to penalizing female candidates and, consequently, presenting male candidates as more viable~\cite{r30}.\par  
Although many requirements engineering (RE) methods exist to capture and analyze functional and non-functional requirements of systems with varying levels of complexity and application domains, with the advancement in technologies and data-driven methods, collecting traditional requirements is no longer sufficient to ensure high quality and sound outputs of systems. It is argued that there is a need to enhance existing RE activities or propose new methods to adapt to the inductive nature of ML requirements, deal with the continuously changing requirements in ML models, and manage various challenges such as performance drift and ethics~\cite{r8,r4,r3,r9}. To ensure robustness, reliability, and fairness in ML systems, new categories of requirements—particularly those related to data and ethics—should be systematically collected and analyzed. Although recent research has addressed the use of goal modeling in ML projects~\cite{Gike2022,6290700, Neace2017, DIMITRAKOPOULOS201928,city23738}, these studies have not addressed capturing data requirements and analyzing alternative datasets. Data requirements define the necessary properties, constraints, and quality attributes of training data of an ML model based on the problem to be solved. These requirements guide the elicitation, specification, and verification of data needed to support the development and deployment of ML-based systems \cite{dey2023multi}. Therefore, the aim of this study is to \emph{propose an approach for identifying and capturing data requirements that must be met prior to using a dataset in the development of ML systems.}\par

To achieve our goal, we adopted Goal-oriented Requirement Language (GRL) because of its effectiveness to represent objectives, constraints, and trade-offs capability. GRL would allow the requirements engineers along with the data scientists to link the data requirements to higher-level system goals hence ensuring that critical data properties are systematically identified and analyzed in the context of the ML-based system’s success criteria. Moreover, GRL was selected because it enables the evaluation of alternative datasets using formal evaluation strategies to see which dataset better satisfies the defined ML-based system's goals.

Therefore, we conducted a literature review to identify the key data requirements necessary for building the Data Requirements Goal Model (DRGM). We then developed a customization mechanism for the DRGM to adapt to various ML problems and contexts. This customization mechanism was designed based on insights from both gray literature and scientific literature, as well as expert input, and includes two UML activity diagrams that assist requirements engineers in the customization process. The resulting DRGM identifies essential data requirements for ML systems and supports requirements engineers in evaluating dataset alternatives. The effectiveness of the proposed DRGM and customization mechanism was assessed through two case studies in different contexts. This work is expected to benefit requirements engineers, especially those with limited ML expertise, by aiding in data planning and ensuring that datasets meet critical requirements. The key contributions of this work are as follows:
\begin{enumerate}
    \item Propose the use of goal-oriented modeling to capture key data requirements for developing ML systems.
    \item Develop a customization mechanism to adapt the data requirements goal model to various problems and contexts.
    \item Evaluate the proposed approach using two illustrative examples based on real-world projects.
\end{enumerate}

This paper is organized as follows. Section~\ref{section2} presents the literature review, addressing the challenges encountered in developing data requirements for ML systems. Section~\ref{section3} details the methodology of this study, including the construction of the DRGM and the customization mechanism, along with illustrative examples that demonstrate their effectiveness. Finally, Section~\ref{conc} presents the conclusion, discusses study limitations, and offers suggestions for future work.\par

\section{Related Work}
\label{section2}
Recent literature reviews emphasized the importance of requirements processes for eliciting requirements specific to ML systems~\cite{habiba2024mature,Khloodslr}. Belani et al. proposed the RE4AI taxonomy, which mapped ML system challenges related to data, models, and systems to various RE processes from a requirements engineering perspective~\cite{r1}. Cerqueira et al.~\cite{Cerqueira2022guide} presented the RE4AI framework, which integrated ethical considerations into the requirements elicitation process for AI systems. Their study emphasized the lack of adequate training in AI ethics among software development teams and highlighted the need for greater focus on ethics throughout the software development phases. Silva et al.~\cite{de2022requirements} focused on technical aspects of requirements elicitation for AI, outlining tools and techniques essential for capturing AI-specific data requirements. Their findings stressed the importance of adaptability in requirements models to accommodate diverse data attributes. Vogelsang and Borg~\cite{r2} and Horkoff et al.~\cite{r8} agreed that ML-based systems had shifted the development paradigm from coding to training, suggesting that RE practices needed to evolve to address this shift~\cite{r2}.

To address the importance of capturing AI-specific requirements, recent work~\cite{Gike2022,r5,6290700, Neace2017, DIMITRAKOPOULOS201928,city23738} highlighted the role of goal-oriented modeling in refining requirements elicitation and analysis for ML systems and in addressing the unique challenges posed by ML requirements. These studies, however, revealed notable differences in focus areas and methodologies. For instance, $i^*$~\cite{yu2010social} was applied to elicit requirements and model concepts related to AI applications for individuals with dementia~\cite{city23738} and for rehabilitation care~\cite{Gike2022}. FLAGS~\cite{10.1109/RE.2010.25} was employed in~\cite{6290700} to elicit requirements for a surgical robotic assistant, with the resulting goal model subsequently converted to UML to support the development process. As for analysis, Ishikawa and Matsuno~\cite{r5} presented GORE-MLOps to capture the uncertainty and unpredictability inherent in ML systems during implementation. GORE-MLOps modeled different scenarios to meet top-level goals by extending GRL to include three states: feasibility unknown, feasibility validated, and feasibility invalidated. Initially, each goal was assigned the status of feasibility unknown, which was then updated to feasibility validated or feasibility invalidated based on experimental or implementation results. The study introduced the terms proved, denied, or unproved at the leaves of the goal model tree, indicating whether the initial contribution scores assigned to goals were confirmed by experiments. The method demonstrated its effectiveness through an illustrative example.

One of the most recent works in this area was presented by Barrera et al.~\cite{10.1016/j.csi.2023.103806}. The work extended $i^*$ to capture ML requirements and presented a metamodel that included ML concepts such as MLGoals, MLTask, Indicator, Dataset, and MLQualityAspects. This metamodel was constructed by the authors and used through a requirements questionnaire and in collaboration with an ML expert to ensure coverage of all relevant aspects of the ML solution. The metamodel was validated through two use cases: one in an industrial context and another in healthcare. In both cases, the metamodel significantly contributed to exploring project goals, identifying key non-functional requirements, specifying required dataset attributes, and recommending algorithms within a defined subset. Despite the range of work that addressed goal-oriented modeling for eliciting and analyzing requirements for ML-based systems, there is a lack of studies focusing on data requirements in terms of elicitation, evaluation of alternatives, or trade-offs. Thus, this work aimed to bridge this gap by providing an approach for eliciting data requirements and analyzing alternative datasets.

\section{Research Methodology}
\label{section3}
\begin{figure}[h!]
 \centering
 \includegraphics[width=\columnwidth]{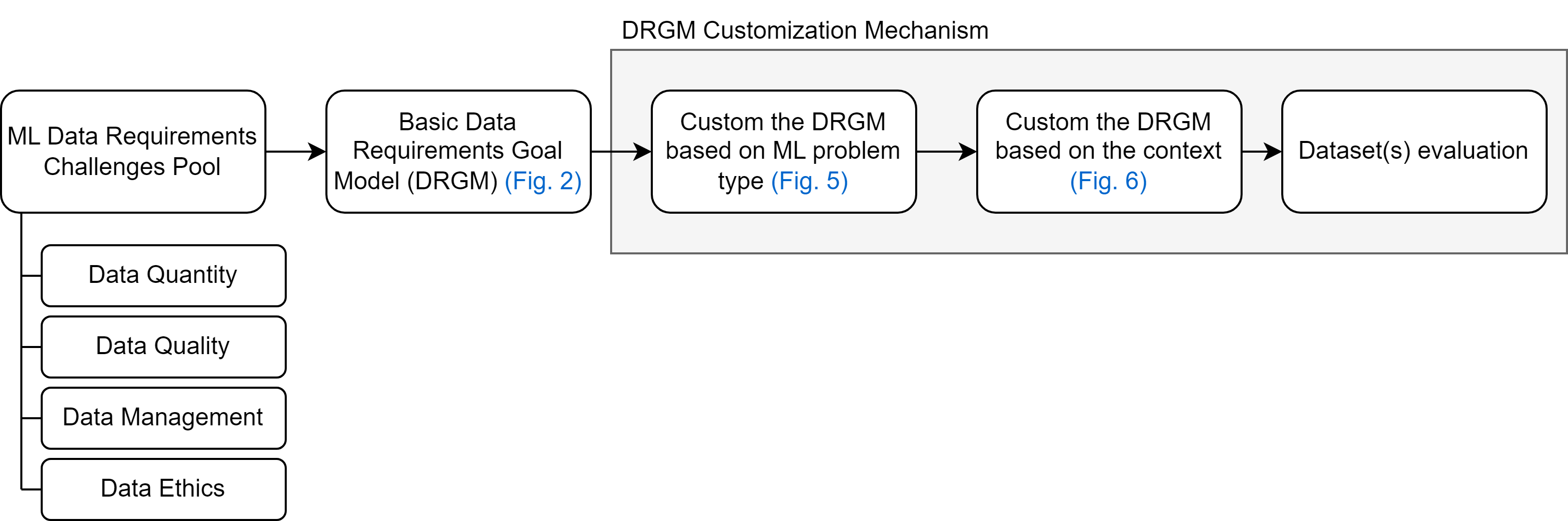}
 \caption{Overview of the Research Methodology}
 \label{FConcept}
\end{figure}

This section presents the DRGM and how it was constructed. It also discusses how the DRGM can be customized to suit different ML problem types and contexts. An overview of the method is presented in Fig.~\ref{FConcept}.

\subsection{Data Requirements Challenges}
\label{sec:2}
 We first identify the key data challenges that requirements engineers may encounter when working on ML systems through a conventional literature review. These challenges were then mapped into four primary categories representing data properties, as outlined by Amershi et al.~\cite{r11}: data quantity, data quality, data management, and data ethics. Table~\ref{table1} provides a summary of the identified challenges, along with their descriptions and corresponding categories.
\begin{table*}[h!]
\centering
\caption{Data Requirements Challenges\label{table1}}
\resizebox{1\textwidth}{!}{
\begin{tabular}{p{0.15\textwidth} p{0.15\textwidth} p{0.5\textwidth} p{0.2\textwidth}}
\hline
\textbf{Category} & \textbf{Challenge} & \textbf{Description} & \textbf{Study} \\
\hline

\multirow{2}{*}{Data Quantity}
  & Data availability
  & The existence of the data
  & \cite{r11,r1,r13,r8,9796389} \\
  & Data accessibility
  & The data can be reached
  & \cite{r11} \\
\hline

\multirow{6}{*}{Data Quality}
  & Data accuracy
  & Data values are correct
  & \cite{r11,r6,r2} \\
  & Data freshness
  & Concerns the latency when receiving the data, and how long it takes to process incoming data
  & \cite{r11,9796389} \\
  & Data representativeness
  & Assumes that the dataset covers the investigated population by having sufficient and similar distribution in the training and testing datasets
  & \cite{r6,9796389} \\
  & Data balancedness
  & The number of samples per category should be balanced
  & \cite{r1,r6,9796389} \\
  & Data completeness
  & Data covers all possible values of the context
  & \cite{r6,r2} \\
  & Data consistency
  & All data in the dataset should be in the same format and representation
  & \cite{r2} \\
\hline

\multirow{2}{*}{Data Management}
  & Data logging
  & Collecting and saving data over time while keeping track of its metadata
  & \cite{r11,r3} \\
  & Data security
  & Confidentiality and integrity of the data
  & \cite{r8,r4} \\
\hline

\multirow{4}{*}{Data Ethics}
  & Data discrimination
  & Presence of protected attributes (or proxies) or data biased toward a certain group
  & \cite{r2} \\
  & Data legality
  & Constraints on obtaining and using the dataset
  & \cite{r2,9796389} \\
  & Data privacy
  & Data should not be shared or used for other purposes, especially personal data
  & \cite{r1,9796389,r8,r4,r7} \\
  & Data safety
  & Protecting data from risk or uncertainty and preserving sensitive data
  & \cite{r1,r4,r7} \\
\hline

\end{tabular}
}
\end{table*}

\subsection{Data Requirement Goal Model (DRGM)}

\begin{figure}[h!]
  \centering
  \graphicspath{{./images/}}
  \includegraphics[width=\columnwidth]{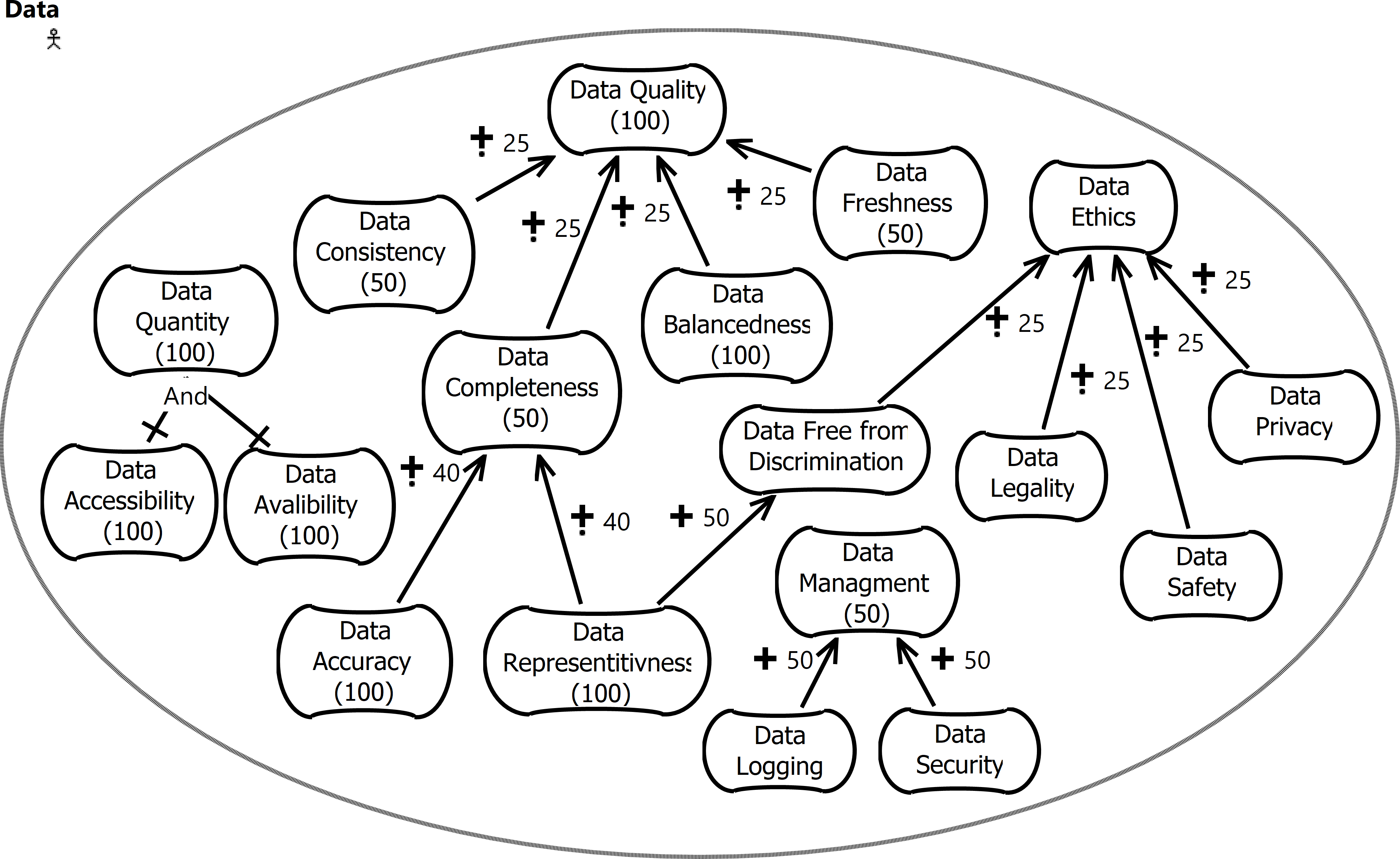}
  \caption{Original Data Requirements Goal Model}
  \label{F1}
\end{figure}

\begin{figure*}[h!]
  \centering
  \graphicspath{{./images/}}
  \includegraphics[width=0.7\textwidth]{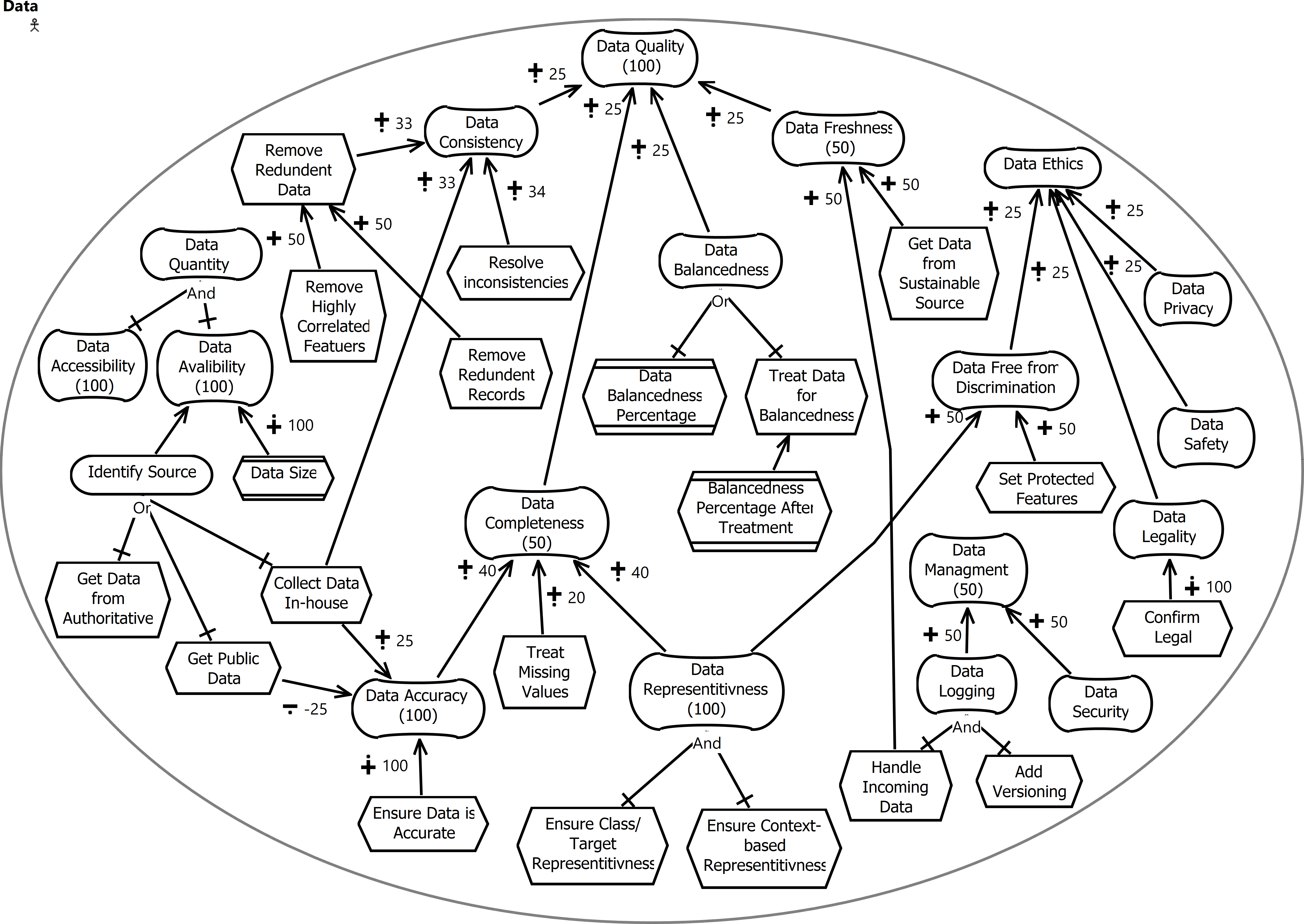}
  \caption{Operationalization of the Data Requirements Goal Model}
  \label{F2}
\end{figure*}

The Data Requirements Goal Model (DRGM), shown in Fig.~\ref{F1}, was constructed by mapping the data requirements challenges listed in Table~\ref{table1} to GRL elements. We used soft goals to capture the essential data properties and qualities required when preparing datasets for ML systems. Those soft goals are to satisfy the data requirements for a ML-based system, which are encapsulated in the data actor that will be the basis for our evaluation. Additionally, suggested methods to handle these data requirements challenges were captured from gray and scientific literature. Authors were reporting them as part of the preprocessing steps or as recommendations to enhance a ML model's performance. The relationships between the DRGM elements are illustrated using contribution and decomposition links in Fig.~\ref{F2}\footnote{Higher resolution images are in: https://github.com/asmayamani/DRGM}. Initial importance values were assigned to the DRGM elements subjectively by the authors after surveying the literature, and these values will be reassigned using the customization mechanism based on the ML problem type and context.


\textbf{Data Quantity} is composed of two subgoals: \emph{Data Availability} and \emph{Data Accessibility}. The \emph{Data Availability} goal was further decomposed into a subgoal for identifying the data source and a KPI element for data size, with a 'make' contribution to ensure the goal’s satisfaction. Having a large dataset when training a model from scratch is essential. A small dataset may amplify outliers and fail to capture the full variance in the sample space~\cite{r2}. The \emph{Data Accessibility} is the second decomposition of the \emph{Data Availability} as the availability of the data does not ensure that it is usable, reachable, or understandable. The data source also impacts data quality goals. For example, using public data may compromise \emph{Data Accuracy}, as public datasets can contain labels from non-experts, potentially leading to incorrect data~\cite{r2}. Conversely, collecting data in-house can enhance data accuracy and consistency~\cite{r2,r11}. In certain contexts, obtaining data from an authoritative source is also crucial~\cite{r11}. Therefore, it is a critical task to ensure that data is accurate.

\textbf{Data Quality} consists of four subgoals: \emph{Data Consistency}, \emph{Data Completeness}, \emph{Data Balancedness}, and \emph{Data Freshness}. \emph{Data Consistency} can be achieved through data preprocessing such as resolving inconsistencies to refine the dataset~\cite{r2} and by removing redundant data. For \emph{Data Balancedness}, users must first evaluate whether the dataset is balanced with respect to the contributing KPI. If imbalances are found, the dataset should be treated for balancedness before assigning the KPI percentage. \emph{Data Completeness} is achieved when the data represents the entire range of possible values relevant to the problem. The subgoal \emph{Data Accuracy} also contributes to it as inaccurate values can render the data incomplete, and a task to 'Treat Missing Data', which can involve interpolation or other techniques. For \emph{Data Representativeness}, the model should be built and tested on data that represents the target phenomena with a similar data distribution. It is essential to ensure that the dataset includes all target classes or value ranges. Additionally, context-based representativeness such as spatial, seasonal, or other domain-specific factors must be ensured, as a lack of representativeness could lead to data discrimination~\cite{r2,r6}. \emph{Data Freshness} is the final component of data quality. Its importance varies depending on the context, which will be discussed later. Two tasks contribute to \emph{Data Freshness}: handling incoming data correctly and obtaining data from sustainable sources~\cite{r11}. 

\textbf{Data Management} comprises two subgoals: \emph{Data Security} and \emph{Data Logging}. \emph{Data Logging} is crucial in many ML applications where a stream of data is available to improve the model. Therefore, this requires a method to handle and log incoming data with versioning the data so that new data is not mixed with already trained data~\cite{r11}. \emph{Data Security} is equally important to protect and preserve the confidentiality, integrity, and availability of data.
\textbf{Data Ethics} consists of four subgoals: \emph{Data Legality}, \emph{Data Privacy}, \emph{Data Safety}, and \emph{Data Free from Discrimination}. \emph{Data Legality} is critical in specific contexts, such as finance, where certain domains have strict legal requirements~\cite{r2}. Consequently, a task to confirm compliance with context-specific legal constraints was added to the \emph{Data Legality} soft goal. Ensuring data is free from discrimination is vital, as biases against specific groups can propagate through the model, causing fairness issues~\cite{10.1145/3468264.3468536}. It is therefore essential to identify features that could lead to discrimination and protect them, such as gender~\cite{r2}. Nonetheless, \emph{Data Privacy} is critical data ethics concern that ensure individuals' cannot be identified. Moreover, protecting them from corruption and ensuring their validity are ethical aspects represented in the \emph{Data Safety}.

\subsection{Customization Mechanism of the DRGM}
\label{section3.5}
\begin{figure}[h!]
  \centering
  \graphicspath{{./images/}}
  \includegraphics[width=0.6\columnwidth]{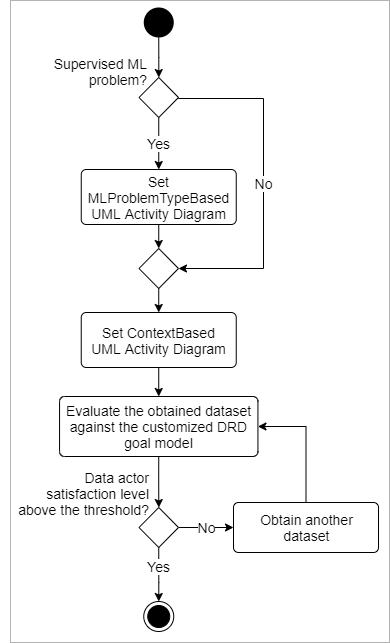}
  \caption{The Customization Mechanism}
  \label{F13}
\end{figure}

The customization mechanism is designed to adapt the initial DRGM to different ML problems and contexts. In addition, it assists requirements engineers, particularly those who are non-experts in ML, in selecting between datasets or evaluating a dataset to ensure that it satisfies the data requirements for a given ML problem, as illustrated in Fig.~\ref{F13}. \par


Given that data requirements are highly dependent on the type of ML problem and are often context-specific, the customization mechanism categorizes data requirements goals into three sets. The first set includes goals that are of high importance across all ML problems, referred to as goal set \#1. The second set comprises goals and subgoals that vary based on the ML type (e.g., Regression, Classification, Time Series). The third set pertains to data requirements that vary based on the collected data and the specific context. The contents of each set are detailed in Table~\ref{table2}. Tasks can be added as illustrated in Fig \ref{F2}. \par




\begin{table}[h!]
\centering
\caption{The customization mechanism sets}
\label{table2}
\resizebox{\columnwidth}{!}{
\begin{tabular}{lll}
\hline
\textbf{Set Number} & \textbf{Elements} & \textbf{Type} \\
\hline

\multirow{2}{*}{Set \#1}
  & \makecell[l]{Data Quantity\\Data Quality}
  & Main goals \\
  & \makecell[l]{Data Availability\\Data Completeness\\Data Consistency\\
                 Data Safety\\Data Accessibility\\Data Accuracy}
  & Subgoals \\
\hline

\multirow{2}{*}{Set \#2}
  & Data Balancedness
  & Subgoals \\
  & \makecell[l]{KPI on Data Availability\\KPI on Data Balancedness}
  & KPI \\
\hline

\multirow{2}{*}{Set \#3}
  & \makecell[l]{Data Ethics\\Data Maintainability}
  & Main goals \\
  & \makecell[l]{Data Security\\Data Legality\\Data Privacy\\
                 Data Free from Discrimination\\Data Freshness}
  & Subgoals \\
\hline

\end{tabular}
}
\end{table}

To use the customization mechanism, the user begins with the initial DRGM, where the goals in set \#1 are fixed. Next, the user customizes the goal model by incorporating elements that depend on the ML problem type using the \emph{MLProblemTypeBased} UML activity diagram shown in Fig.~\ref{F14}. Afterward, the user employs the \emph{ContextBased} UML activity diagram in Fig.~\ref{F15} to define context-based elements. Once the customization is complete, the user evaluates each dataset using GRL evaluation strategies using the customized goal model. If the evaluation indicates that the \emph{Data Actor} is not satisfied, another dataset should be selected, or additional preprocessing should be applied. This evaluation process is iterative and continues until the \emph{Data Actor} is satisfied. A detailed explanation of how this customization mechanism handles different data requirements sets is provided in the following subsections. \par

\subsubsection{Goals with Fixed Importance}
Examining the factors that determine set \#1, \emph{Data Quantity} is consistently assigned a 'High' importance, as having more high-quality data generally leads to better results~\cite{r2}. To satisfy the \emph{Data Quantity} goal, both \emph{Data Availability} and \emph{Data Accessibility} must also be satisfied. \emph{Data Quality} is equally critical; if \emph{Data Quality} is not met, the acquired data will be ineffective, resulting in a 'garbage-in, garbage-out' scenario, as noted by a data scientist in the study~\cite{r2}. Similarly, \emph{Data Representativeness}, \emph{Data Completeness}, and \emph{Data Accuracy} hold high importance across all contexts. That is, for \emph{Data Representativeness}, a dataset must reflect the underlying population or environment it aims to model. Lack of representativeness leads to sampling bias, which causes poor generalization and fairness issues in downstream models \cite{torralba2011unbiased}. Moreover, \emph{Data Completeness} must be ensured to enable robust decision-making as missing values can lead to inaccurate or unstable models \cite{little2019statistical}. Noisy input or incorrect labels have been shown to reduce accuracy and affect model convergence \cite{frenay2013classification}, hence, \emph{Data Accuracy} holds high importance as well. Although \emph{Data Consistency} and \emph{Data Safety} are also important, their significance is comparatively lower, and their importance is therefore set to 'Medium'. This is because they are often context-dependent \cite{batini2016data}. That is, they are critical in regulated or adversarial settings but typically of medium importance in performance evaluation without such constraints \cite{batini2016data}.\par

\subsubsection{Goals with Different Importance Based on ML Problem Type} 
\label{MLProblemType} 
Two elements related to data requirements vary in importance depending on the type of ML problem: \emph{Data Balancedness} and \emph{Data Availability}. \emph{Data Balancedness} is highly important for classification problems but holds moderate importance for regression problems. For classification problems, techniques such as SMOTE, oversampling, and undersampling can be applied to address imbalanced data~\cite{r15,r14}. For regression problems, data transformations to handle imbalance, often referred to as skewness, can be used~\cite{r16,r17}. To set KPIs for \emph{Data Balancedness} in classification problems, users should evaluate the balance as a percentage from the ratio (e.g., for binary classification, a 1:1 class ratio corresponds to 100\% balance). For regression problems, the Shapiro-Wilks test is suggested to detect imbalances \cite{mishra2019descriptive}. This test's null hypothesis states that the data is drawn from a normally distributed population. If \( P > 0.05 \), the null hypothesis is accepted, and the data is considered normally distributed. If the null hypothesis is rejected, the target variable is highly skewed~\cite{r18}. The valuation values are determined based on the percentages or test results.

For setting KPIs for \emph{Data Availability}, several approaches exist to calculate the required amount of data based on the learning rate~\cite{r19}. However, these approaches require an initial dataset and a pre-built ML algorithm. As such, they are primarily useful for determining the additional data needed during a pilot study for a selected dataset. When no initial dataset exists, there are no solid, theoretically proven rules for determining how much data an ML algorithm requires for training. Nonetheless, a heuristic known as the 'rule of 10,' based on practitioners’ experiences, is frequently mentioned in the gray literature~\cite{r21,r22,r20} and in a white paper for classification problems~\cite{r23}. This rule suggests using ten instances per class or ten instances per predictor. Variations of this rule propose reducing the requirement to five instances or increasing it to 100. For this work, the KPI is set following this rule, incorporating its three variations which are classification, regression, and time series. 

\begin{figure*}[h!]
  \centering
  \graphicspath{{./images/}}
  \includegraphics[width=0.8\textwidth]{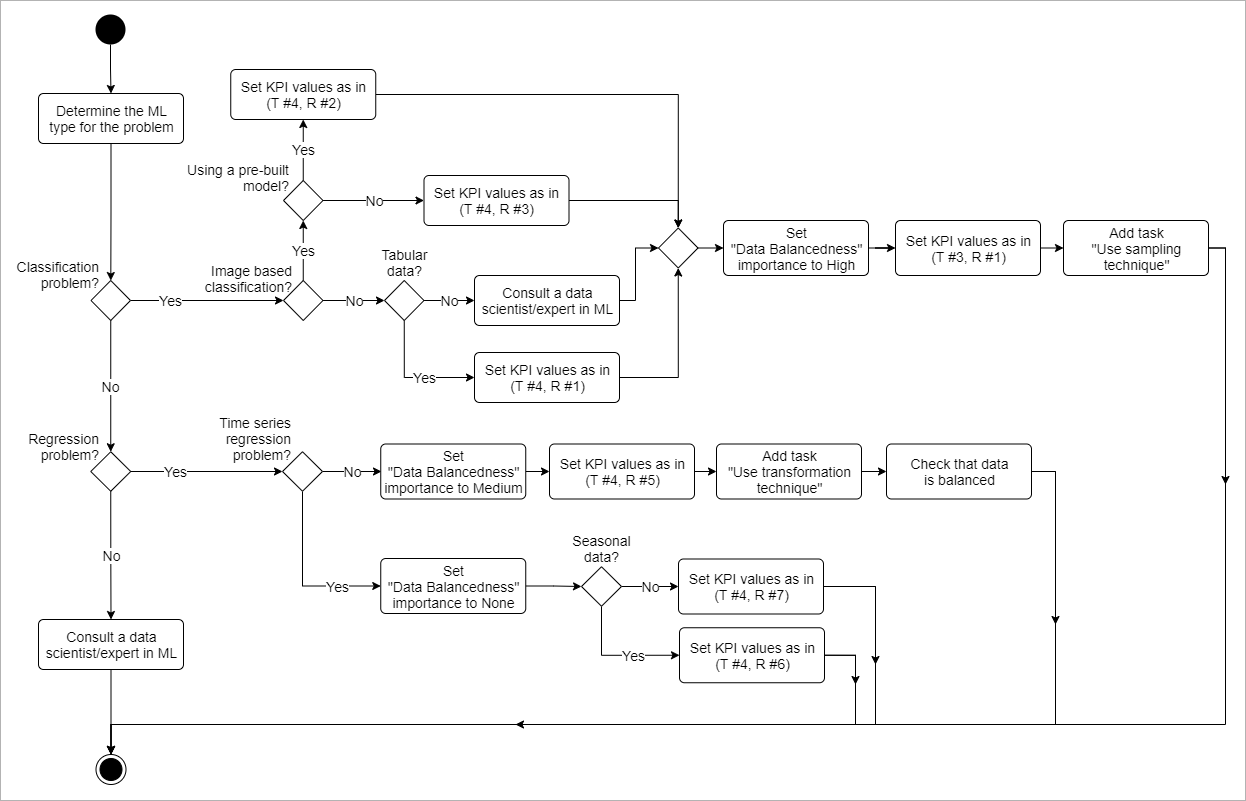}
  \caption{ML Problem Type Based UML Activity Diagram}
  \label{F14}
\end{figure*}

In the case of image classification using deep learning, if a pre-trained model is not used, 1000 images per class are typically required as the minimum threshold~\cite{r24}. Additional KPI values are derived from practitioners' experiences and estimates. For forecasting with time-series data, particularly in seasonal data (e.g., weather or sales data), authors in \cite{r25} suggest using the number of seasons within a year plus five additional data points. However, real-world scenarios often involve randomness, requiring the model to utilize multiple 'seasons' worth of data. Based on~\cite{9718423}, the worst-case requirement is set to one times the number of data points in a season, the threshold to two times, and the maximum to ten times the number of data points in a season. For univariate, non-seasonal data using statistical models, the latest 50 observations are generally sufficient for short-term forecasts~\cite{r27,r28}. Some practitioners from the gray literature suggest $40$ observations may suffice, while others recommend using up to $100$ for more accurate results. For multivariate time-series predictions, the 'rule of 10' is also applied to refine the estimate. A detailed explanation of how to set the data size KPI for various ML problem types is provided in Table~\ref{tab:kpiValuesDataSize}, and the process is summarized in the \emph{MLProblemTypeBased} UML activity diagram shown in Fig.~\ref{F14}.


\begin{table*}[htbp]
\centering
\caption{KPI values for data size}
\label{tab:kpiValuesDataSize}
\resizebox{\textwidth}{!}{%
\begin{tabular}{p{0.2\textwidth}p{0.2\textwidth}p{0.2\textwidth}p{0.2\textwidth}p{0.2\textwidth}}
\toprule
\textbf{ML Problem Type - More specifications} & \textbf{Worst Value} & \textbf{Threshold Value} & \textbf{Target Value} & \textbf{Unit} \\ 
\midrule
Classification - Tabular & Max (5 * \#classes, 5 * \#features) & Max (10 * \#classes, 10 * \#features) & Max (100 * \#classes, 100 * \#features) & Data points \\[1ex]
Classification – Image  & 500 * \#classes & 1000 * \#classes & 10,000 * \#classes & Number of images \\[1ex]
Classification - Other & \multicolumn{4}{c}{Consult a Data Science Expert in the specific domain} \\[1ex]
Regression & 5 * \#features & 10 * \#features & 100 * \#features & Data points \\[1ex]
Time Series - Seasonal & 1 & 2 & 10 & \begin{tabular}[c]{@{}l@{}}Years (or the maximum \\ seasonality period) worth of data\end{tabular} \\[1ex]
Time Series - Other & Max (40, 5 * \#features) & Max (50, 10 * \#features) & Max (100, 100 * \#features) & Data points \\ 
\bottomrule
\end{tabular}%
}
\end{table*}

\subsubsection{Goals with Different Importance Related to Context}
Many data requirements are context-dependent. For the \emph{Data Representativeness} goal, the dataset should adequately capture variations across different geographic regions (spatial data) and time periods or seasons (seasonal-temporal data), to ensure the model can generalize across diverse conditions. If the prediction outcome directly impacts human subjects (e.g., in healthcare, criminal justice, or employment), it is critical to ensure \emph{Data Representativeness} across targeted demographics. Lack of representativeness can lead to biased model predictions, resulting in discriminatory outcomes for underrepresented groups.\par
Regarding \emph{Data Management}, its importance is set to 'Low' only if the model is built once and never updated. In such cases, \emph{Data Freshness} is also set to 'Low'. However, a study by~\cite{r32} found that one-third of models require updates at least monthly, and nearly one-quarter require daily tuning. This is especially relevant in fields such as marketing, stock market forecasting, and short-term weather prediction. For models requiring regular updates, \emph{Data Management} is assigned 'High' importance. Similarly, \emph{Data Freshness} is set to 'High' for models requiring frequent tuning, and the data must be acquired from sustainable sources that are ethically, legally, and practically viable for long-term use in ML without violating privacy, exhausting resources, or causing harm. For models with irregular updates, the importance of \emph{Data Freshness} and sourcing sustainable data is set to 'Medium'. \par

When considering \emph{Data Ethics}, the content of the dataset must first be evaluated. If the data identifies human subjects, identifying information must be removed or protected to prevent discrimination and ensure the privacy. Additionally, user consent must be obtained as a task under \emph{Data Privacy}. GDPR regulations impose stricter consent requirements for identifying information, such as a photograph of a person’s house. These rules apply when the subjects are European Union (EU) citizens or if the system will be used within the EU~\cite{r29}. In other regions, compliance should align with the relevant local regulations.\par

\begin{figure*}[h!]
  \centering
  \graphicspath{{./images/}}
  \includegraphics[width=0.8\textwidth]{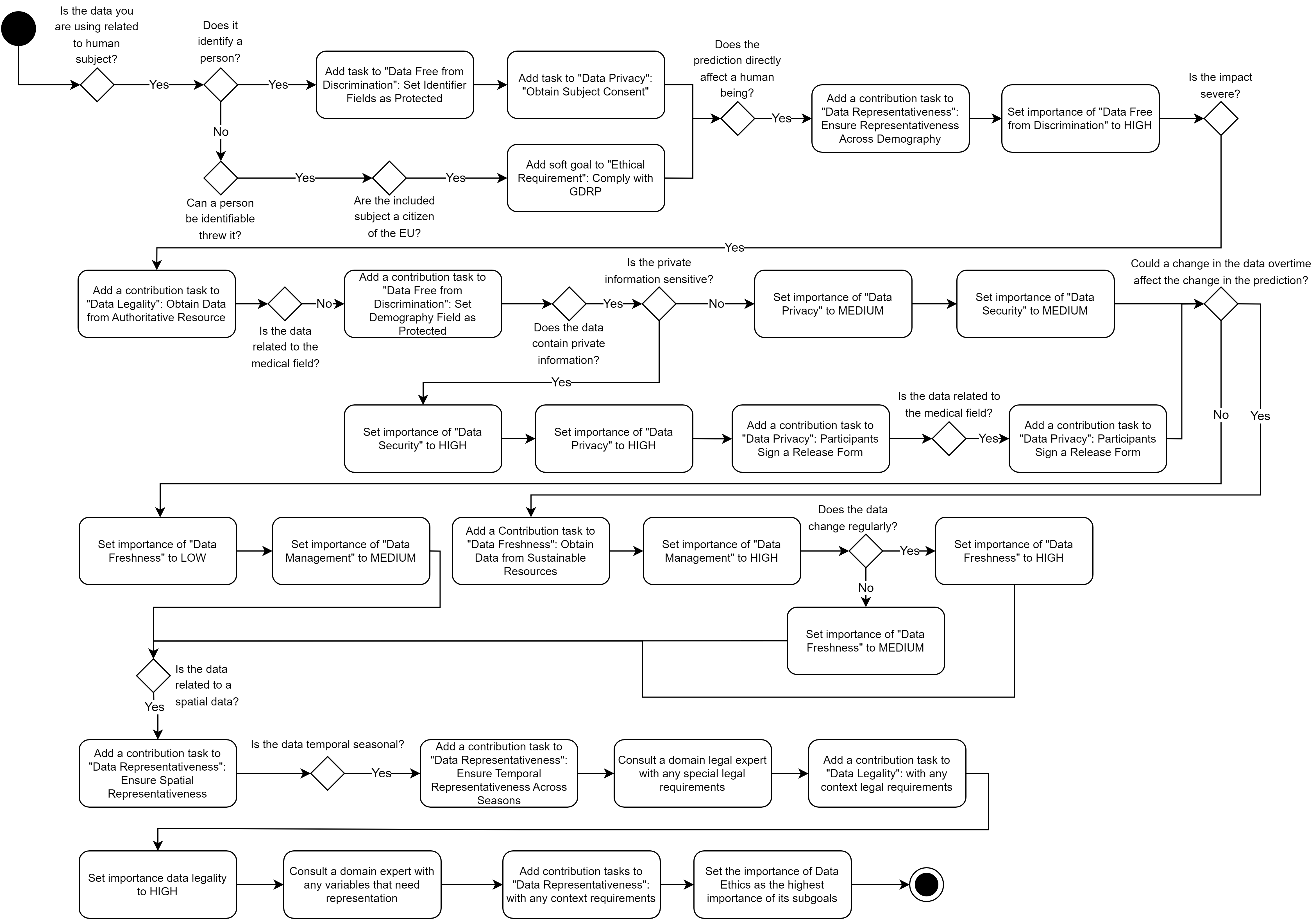}
  \caption{Context Based Activity Diagram}
  \label{F15}
\end{figure*}

Next, we examine the \emph{Data Privacy} goal. If the dataset contains sensitive information, such as health records, financial records, university records, or business records, the importance of both \emph{Data Privacy} and \emph{Data Security} is set to 'High'. If the data is private but not sensitive, their importance is set to 'Medium'. For public data, the importance of \emph{Data Privacy} and \emph{Data Security} is set to 'None' as not all public data is directly usable. For private data, a release form must be signed to obtain access. For medical data, approval from an International Review Board (IRB) is required. Additionally, the importance of the goal \emph{Data Free from Discrimination} is set to 'High' if the ML model's decisions have a significant impact on human lives, such as in job recruitment~\cite{r30} or prison release~\cite{r31}. In such cases, data should be obtained from authoritative sources, and sensitive demographic fields should either be removed or designated as protected to prevent bias in the system. Finally, the overall importance of \emph{Data Ethics} is determined by the highest importance level among its subgoals. Before initiating the evaluation, an analyst should consult with both domain and legal experts to address any potential misrepresentation in the data and any legal concerns specific to the field. A summary of this process is illustrated in the ContextBased UML activity diagram, Fig.~\ref{F15}. \par

\subsubsection{Dataset Evaluation and Selection}
After customizing the DRGM based on the problem type and context, the dataset(s) are evaluated using the GRL evaluation strategy to select the dataset that best satisfies the \emph{Data Actor}. We propose an evaluation scale ranging from 0 to 100. During the evaluation, all unrelated leaf elements should have their qualitative evaluation values set to 'satisfied,' and their negative contribution links removed to avoid negatively impacting the overall evaluation. 

The data size KPI value should be assigned only after any preprocessing steps that involve the removal of data points have been completed. The selected dataset must satisfy the \emph{Data Actor} with a score above a certain threshold; we propose a minimum threshold of $70$. If multiple datasets meet this threshold, the dataset with the highest satisfaction score should be selected. If no dataset satisfies the \emph{Data Actor}, additional preprocessing may be applied, or alternative datasets may need to be obtained or combined.

\section{Illustrative Examples for Using DRGM}
\label{section4}
This section shows the customization of the DRGM approach using the proposed customization mechanism. Since the data requirements depend on the ML prediction type and are mostly context-related, we choose two different illustrative examples. The first one, in the medical field, uses classification as the ML type. The second example uses the regression for time series weather forecasting. \par 
\begin{figure*}[h!]
  \centering
  \graphicspath{{./images/}}
  \includegraphics[width=0.65\textwidth]{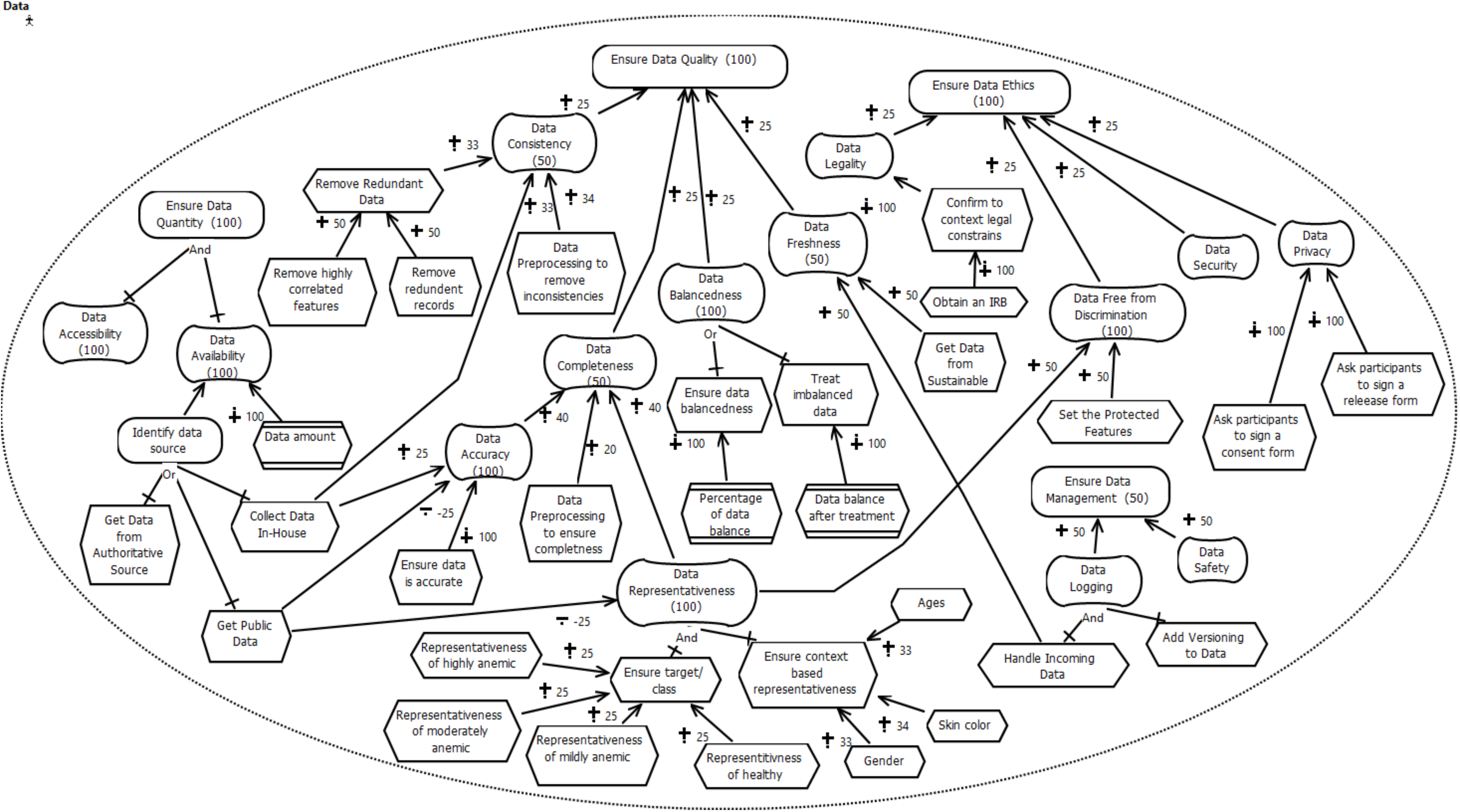}
  \caption{Anemia Detection Customized DRGM}
  \label{F18}
\end{figure*}

\begin{figure}[h!]
  \centering
  \graphicspath{{./images/}}
  \includegraphics[width=\columnwidth]{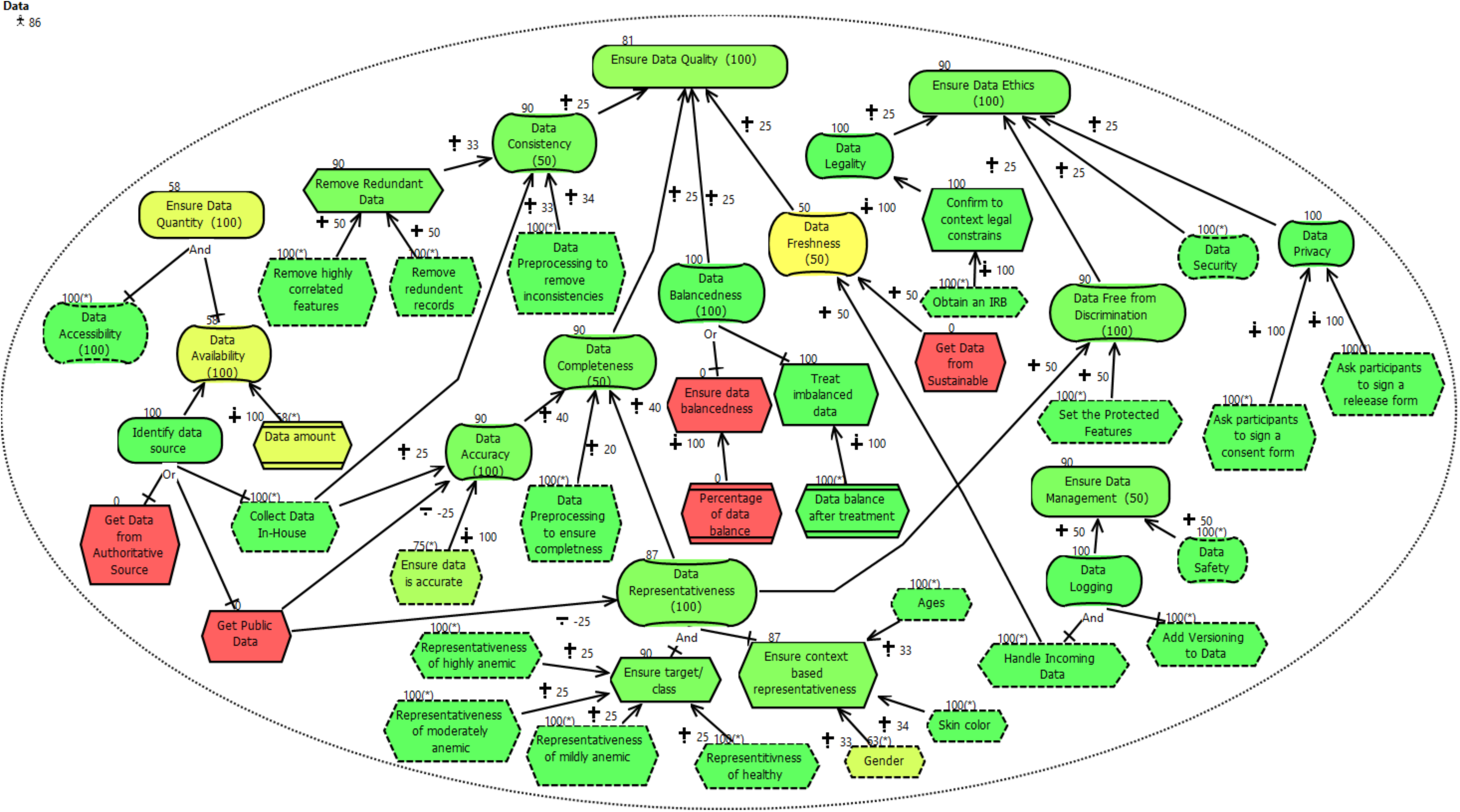}
  \caption{Anemia Detection Dataset\#1 Evaluation Model}
  \label{F19}
\end{figure}

\begin{figure}[h!]
  \centering
  \graphicspath{{./images/}}
  \includegraphics[width=\columnwidth]{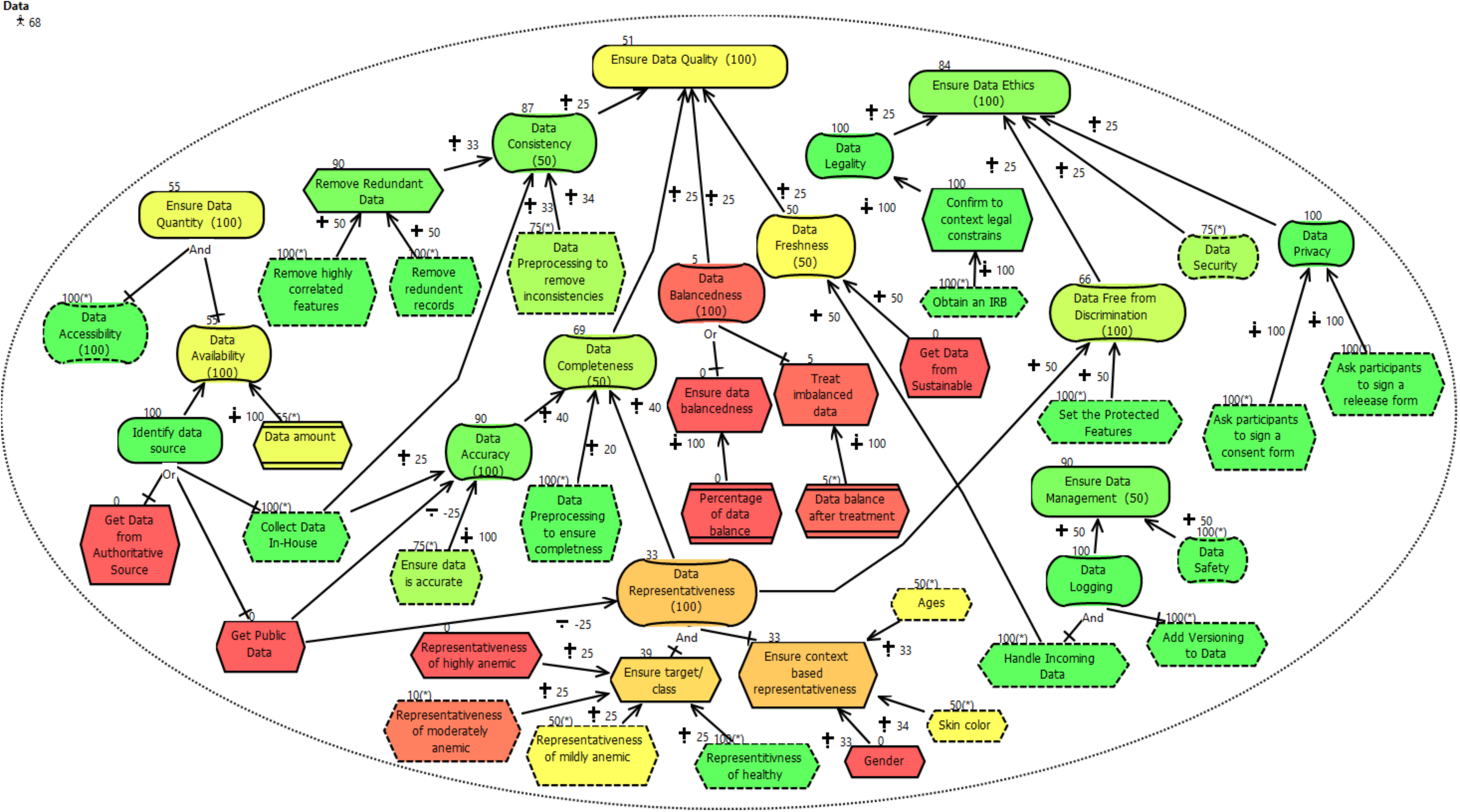}
  \caption{Anemia Detection Dataset\#2 Evaluation Model}
  \label{F20}
\end{figure}

\subsection{Non-invasive Diagnosis of Anemia System using Machine Learning}
\textbf{Context.} Researchers in~\cite{r33} aimed to build an ML system to diagnose anemia using a video of a patient’s fingertip. To collect the training and testing dataset, three infrared LEDs and one white LED were positioned around the camera capturing the video. The LEDs were activated sequentially, and the average values of each of the three channels (RGB) from the video were calculated separately. These values were then used to train an ML model with the target classes being severely anemic, moderately anemic, mildly anemic, or healthy in a multi-class classification problem.\\ 
\textbf{Customizing the DRGM.} First, we follow the \emph{ML problem type based UML activity diagram} and as we are dealing with a classification problem we set the balancedness to 'High', added the KPI as per the diagram, then added the related task. As for the KPIs for the data availability, we have $4$ classes, so we set the worst value to $20$, threshold value to $40$, and target value to $400$. We then follow the \emph{context-based UML activity diagram}, and as the data relates to human subject Data Ethics requirements will be set to 'High'. No special legal requirements other than the IRB and the consent release forms are required. Different age groups, both female and male, and different skin colors should be represented. By completing this step, we now have a customized DRGM based on the ML type and based on the context, Fig.~\ref{F19}.\\
\textbf{Dataset Evaluation.} The initial plan for the dataset involved collecting data in-house from 100 participants. Due to time constraints, the goal was to create a representative dataset and address any imbalance through sampling. Redundant data would be removed, and since the model was intended to be built only once, data would not be sourced from sustainable resources. Evaluating the planned dataset using the customized DRGM, shown in Fig.~\ref{F19}, indicated promising outcomes if the implementation were executed properly. However, as the implementation of the project described in~\cite{r33} progressed, the total number of participants was reduced to $80$. The conventional participant selection process led to a dataset that lacked adequate representation of individuals with fair skin tones, male participants, and those who were moderately or severely anemic. After performing undersampling on the healthy class, the dataset remained highly imbalanced, with class ratios of 3:5:12:40. Consequently, the balancedness KPI was set to $20$. Certain inconsistencies in the dataset could not be resolved, as recordings were conducted in different rooms. Additionally, some accuracy issues arose due to the fading of lights towards the end of the data collection phase. The presence of multiple participants in the recording rooms further posed risks to data safety. The evaluation of the collected dataset, presented in Fig.~\ref{F20}, highlights why the project was ultimately halted. It faced significant challenges, including low recall and poor accuracy for individuals with fair skin tones.

\begin{figure*}[h!]
  \centering
  \graphicspath{{./images/}}
  \includegraphics[width=0.65\textwidth]{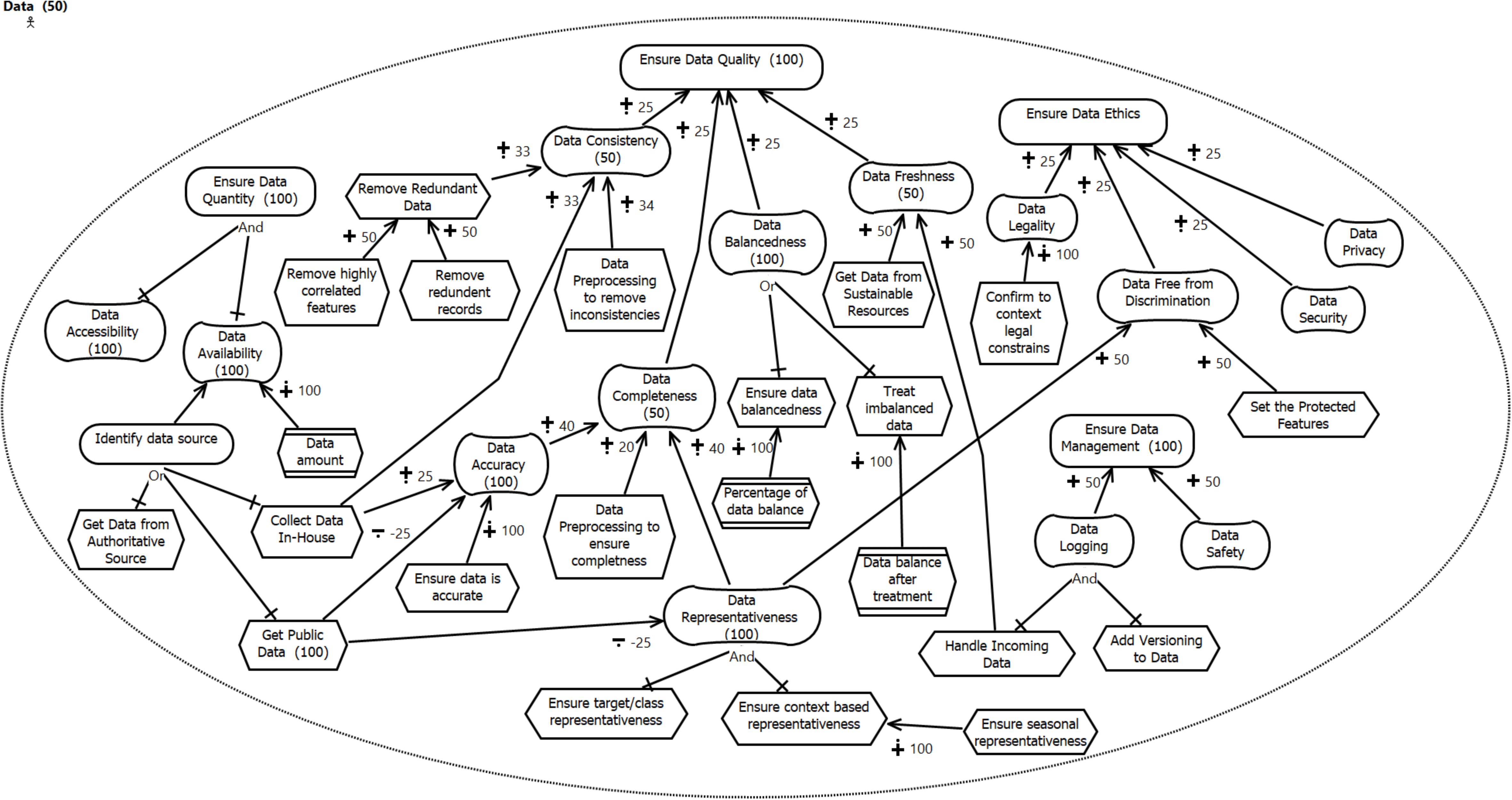}
  \caption{GHI Forecasting Customized DRGM}
  \label{F23}
\end{figure*}

\begin{figure}[h!]
  \centering
  \graphicspath{{./images/}}
  \includegraphics[width=\columnwidth]{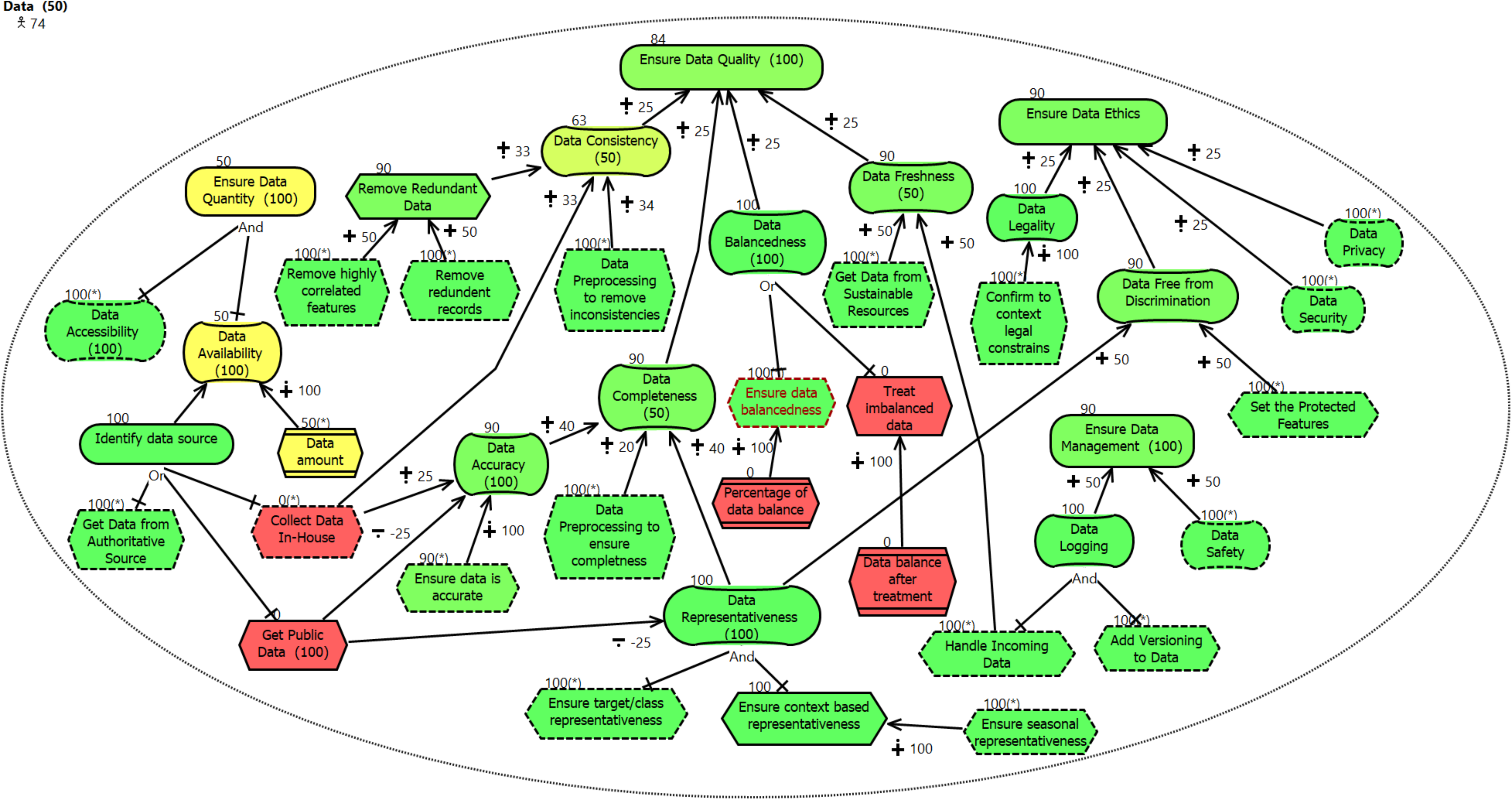}
  \caption{GHI Forecasting Dataset\#1 Evaluation Model}
  \label{F24}
\end{figure}

\begin{figure}[h!]
  \centering
  \graphicspath{{./images/}}
  \includegraphics[width=\columnwidth]{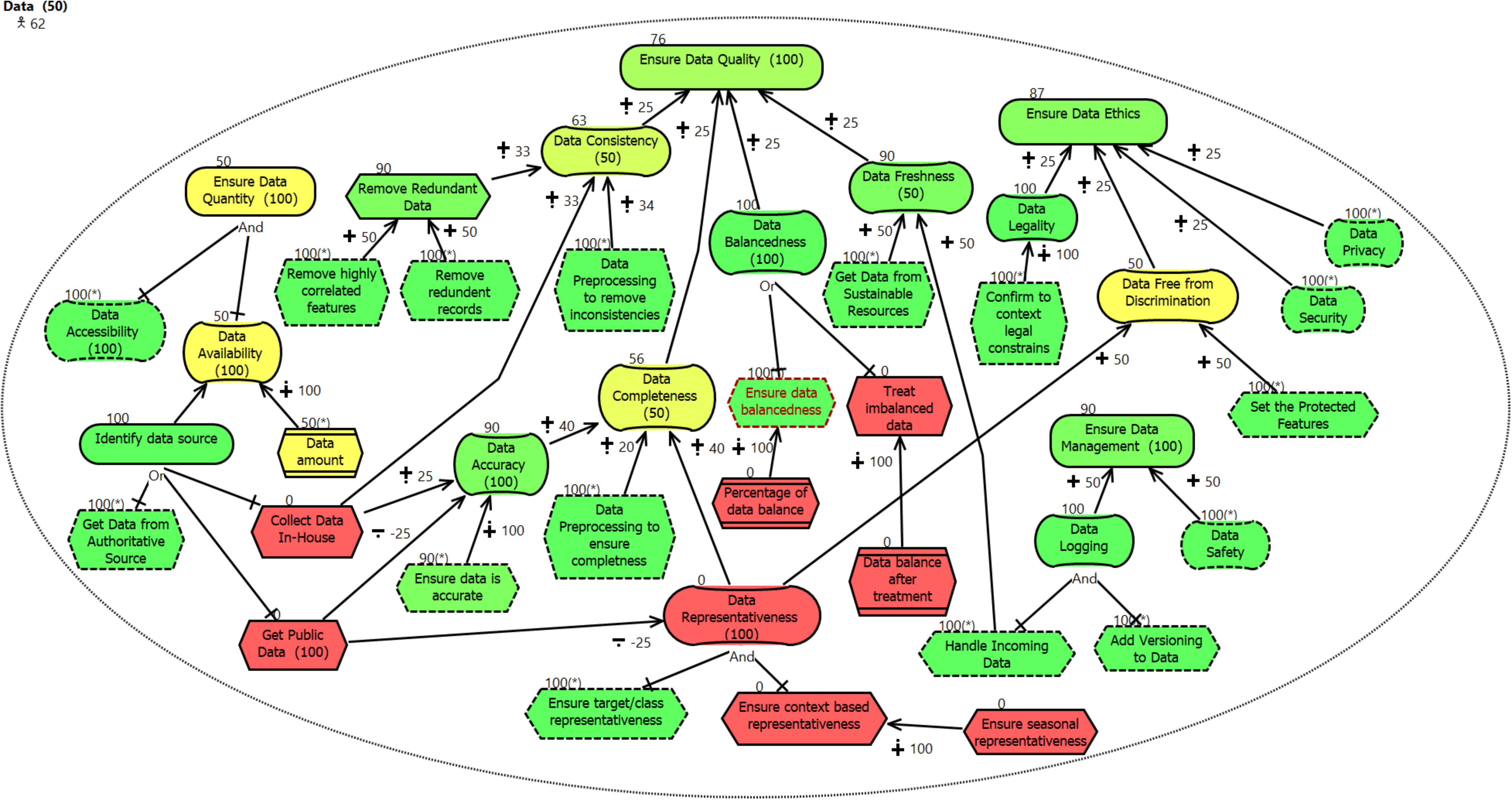}
  \caption{GHI Forecasting Dataset\#2 Evaluation Model}
  \label{F25}
\end{figure}

\subsection{Hourly Global Horizontal Irradiance (GHI) Forecasting}
\textbf{Context.} Researchers in~\cite{9718423} investigated the use of LSTM models to forecast hourly Global Horizontal Irradiance (GHI) for a specific city, a critical attribute in determining the amount of solar energy that can be harvested from solar panel farms.

\textbf{Customizing the DRGM.} First, the \emph{ML Problem Type-Based UML Activity Diagram} was followed. Since this is a time-series regression problem, neither the balancedness of the data nor its redundancy was considered relevant. The amount of data KPI was set as recommended. Next, the \emph{Context-Based UML Activity Diagram} was applied. As the data does not pertain to human subjects, directly impact human subjects, or contain sensitive information, the importance of data ethics requirements was set to 'None'. On the other hand, because the data is time-series in nature, the importance of \emph{Data Freshness} and \emph{Data Management} was set to 'High'. A temporal representativeness task was added to ensure seasonality is adequately captured. The resulting customized DRGM is shown in Fig.~\ref{F23}.\par

\textbf{Dataset Evaluation.} The data used in~\cite{9718423} was obtained from the official website of K.A.CARE~\footnote{https://rratlas.energy.gov.sa/}, the organization responsible for installing and maintaining GHI monitoring stations across Saudi Arabia. This source is considered sustainable, ensures data freshness, and is regarded as authoritative. Incoming data was set to be handled manually. The importance of data ethics requirements was set to 'None' and assigned an initial value of $100$ during the evaluation to prevent negative impacts on the overall assessment. Some inconsistencies were identified but addressed during preprocessing. However, there was a 10\% uncertainty in the measurement instruments, which affected the data's accuracy. As the study in~\cite{9718423} focused on identifying the minimum data period required to achieve excellent performance (characterized by less than 10\% mRMSE), the researchers experimented with various data volumes. They determined that a minimum of two years of data was necessary for representativeness. The dataset evaluation at these values is illustrated in Fig.~\ref{F24}. However, the satisfaction of the \emph{Data Actor} with this limited time frame is contingent on fully capturing temporal fluctuations. When replicating the experiment for another city, Najran, as shown in Fig.~\ref{F25}, the results did not hold. This was due to missing data for several months and the two years of training data not being representative of Najran’s overall climate, as they included atypical precipitation values.

\section{Conclusion and Future work}
\label{conc}
The breakthroughs of ML in real-life applications highlighted the increasing need to address data requirements, as they are an essential building block of data-driven systems. This work aimed to guide non-experts in ML in identifying the essential data requirements based on the ML problem they are attempting to solve. We proposed the use of goal modeling for this task, as it provides an intuitive way to communicate with domain experts and to analyze trade-offs. To construct the Data Requirements Goal Model (DRGM), we surveyed the literature to identify the challenges associated with developing data requirements for ML systems. These challenges were categorized under four themes and presented as soft goals. We then provided a customization mechanism to adapt the model to different contexts and ML problem types, based on insights from the surveyed gray literature and scientific literature. The goal model and its customization mechanism were evaluated using two illustrative examples from two different contexts and ML problem types. The resulting evaluations aligned with the outcomes reported in the respective studies.\par

Regarding limitations, the customization mechanism was designed for the most common types of supervised ML problems, namely regression and classification. In addition to the scientifically scholar publications, some of the KPI values are derived from practitioners' recommendations and gray literature which may introduce subjectivity or lack the rigorous validation typically found in peer-reviewed sources. Users are also required to manually modify the DRGM using the provided flowcharts, which could lead to errors due to misunderstandings, especially among users unfamiliar with GRL. While we attempted to cover as many cases as possible, there remain domain-specific requirements, particularly concerning data legality and data representativeness, that may not be fully addressed. Thus, we do not claim that this model eliminates the need to consult legal or domain experts. However, the model and customization mechanism significantly minimize this need by streamlining the selection process and quickly filtering out inapplicable datasets. Moreover, this work is limited to traditional ML models ; different KPIs, especially related to data quantity, still need to be addressed for pre-trained deep learning models and LLM based applications. \par

For future work, further evaluation of the model on a wider range of ML problems and contexts is recommended. Incorporating evaluations for pre-trained models is also necessary. Additionally, we plan to develop chatbot tool support to automate the customization mechanism, replacing the current manual process. Also, as this work is limited to traditional ML and deep learning models, extra exploration is needed for pre-trained deep learning models and LLM supported models. 

\section*{ACKNOWLEDGMENTS}
The authors acknowledge the support of King Fahd University of Petroleum and Minerals in the development of this work and acknowledge the support provided by the Deanship of Research at King Fahd University of Petroleum \& Minerals (KFUPM). 

\bibliography{conference_101719}
\bibliographystyle{acm}

\end{document}